\documentclass[apj,twocolappendix, numberedappendixn]{emulateapj}
\usepackage[colorlinks,citecolor=blue]{hyperref}
\usepackage{textcomp}
\usepackage{subfigure}
\usepackage{verbatim}
\usepackage{bm}% bold math
\usepackage{hyperref}
\usepackage{amsmath}
\usepackage{graphicx}
\usepackage{epstopdf}
\usepackage{amssymb}
\usepackage{extarrows}
\usepackage{color}
\usepackage{CJK}
\usepackage{cancel}
\usepackage{ulem}
\usepackage[utf8]{inputenc}

\newcommand{\ba}{\begin{eqnarray}}
\newcommand{\ea}{\end{eqnarray}}
\newcommand{\be}{\begin{equation}}
\newcommand{\ee}{\end{equation}}
\newcommand{\gr}{\mathrm{GR}}

\newcommand{\m}{\mathrm{max}}

\newcommand{\au}{\mathrm{AU}}

\newcommand{\IN}{\mathrm{in}}
\newcommand{\OUT}{\mathrm{out}}

\newcommand{\lk}{\mathrm{LK}}

\newcommand{\tot}{\mathrm{tot}}

\def\e1{e_1^2}

\begin{document}
\title{Probing the Spins of Supermassive Black Holes with Gravitational Waves \\from Surrounding Compact Binaries}
\author{Bin Liu$^{1,2}$, Dong Lai$^{2,3}$}
\affil{$^{1}$ Niels Bohr International Academy, Niels Bohr Institute, Blegdamsvej 17, 2100 Copenhagen, Denmark\\
$^{2}$ Cornell Center for Astrophysics and Planetary Science, Department of Astronomy, Cornell University, Ithaca, NY 14853, USA\\
$^{3}$ Tsung-Dao Lee Institute, Shanghai Jiao Tong University, Shanghai 200240, China
}

\begin{abstract}
Merging compact black-hole (BH) binaries are likely to exist in the
nuclear star clusters around supermassive BHs (SMBHs), such as Sgr A$^\ast$.
They may also form in the accretion disks of active galactic
nuclei. Such compact binaries can emit gravitational waves (GWs) in
the low-frequency band ($0.001-1$ Hz) that are detectable by several
planned space-borne GW observatories. We show that the angular momentum vector of
the compact binary ($\textbf{L}_\IN$) may experience significant variation due to the
frame-dragging effect associated with the spin of the SMBH.  The
dynamical behavior of $\textbf{L}_\IN$ can be understood analytically
as a resonance phenomenon. We show that rate of change of
$\textbf{L}_\IN$ encodes the information on the spin of the
SMBH. Therefore detecting GWs from compact binaries around SMBHs,
particularly the modulation of the waveform associated with the
variation of $\textbf{L}_\IN$, can provide a new probe on the spins of SMBHs.
\end{abstract}

%  \rightharpoondown
%\maketitle

\section{Introduction}

The spins of the supermassive black holes (SMBHs) at the centers of galaxies are poorly constrained; this is the
case even for Sgr A$^\ast$ in the Galactic Center \citep[e.g.,][]{Ghez 1998,Ghez 2008,Genzel 2010}.
The spin vector of an accreting SMBH could in principle be constrained by
modeling the accretion/radiation processes
\citep[e.g.,][]{Moscibrodzka 2009,Broderick 2011,Shcherbakov 2012} and comparing with observations, such as those of Sgr A$^\ast$ and M87
from Event Horizon Telescope \citep[e.g.,][]{Dexter 2010,Broderick 2016,EHT 2019}.
The Galactic Center host a population of young massive stars \citep[e.g.,][]{Genzel 2000,Merritt 2013,Alexander 2017};
it has been suggested that
the relativistic frame dragging effect on their orbits could put constraints on the Sgr A$^\ast$'s spin \citep[e.g.,][]{Levin 2003,Fragione 2020}.

Given the fact that the S-stars around Sgr A$^\ast$ are close to the SMBH
\citep[$\sim0.01$pc, with the newly discovered S4714 orbit reaching a pericenter distance of 12.6AU;][]
{S star 1,S star 2,S star 3}, it is likely that binaries of compact objects could be present near SgrA$^\ast$
\citep[e.g.,][]{Antonini 2012,Stephan 2019}.
Similar compact binaries may also exist in nuclear star clusters around other SMBHs
\citep[e.g.,][]{OLeary 2009,Hopman 2009,Leigh 2018}
and/or form in disks of active galactic nuclei \citep[e.g.,][]{McKernan 2012,Bartos 2017,Tagawa 2020}.
These compact binaries may radiate gravitational waves (GWs) in the low-frequency band ($0.001-1$ Hz), which
can be detectable by the planned/conceived space-borne GW observatories \citep[e.g.,][]{Xianyu LISA,Hoang,Deme},
including LISA \citep[e.g.,][]{LISA}, TianQin \citep[e.g.,][]{TianQin}, Taiji \citep[e.g.,][]{TaiJi}, B-DECIGO \citep[e.g.,][]{DECIGO},
Decihertz Observatories \citep[e.g.,][]{DeciHZ}, and TianGO \citep[e.g.,][]{TianGo}.
The orbital parameters for the barycenter motion of the compact binary around the SMBH can be extracted from the GW
signal taking into account the phase change and Doppler shift \citep[e.g.,][]{Inayoshi,Xianyu}.
A recent study \citep[][]{Yuhang PRL} has shown that the orientation change of orbital plane of the compact binary driven by the
gravitational torque from the SMBH can be measurable for sources at distances less than 1 Gpc (depending on the assumed sensitivity of GW detectors).
In this paper, we demonstrate that the
spin of SMBH can significantly modify the orientation dynamics of the compact binary,
even when the binary orbit remains circular (i.e., no Lidov-Kozai oscillations; see below).
Therefore, in principle, detecting such compact binaries in GWs may provide a new
probe to the spins of SMBHs, including that of SgrA$^\ast$.

\section{Compact Binary Near a Spinning SMBH}
We consider a binary with masses $m_1$, $m_2$, semimajor axis $a_\IN$ and eccentricity $e_\IN$,
moving around a SMBH tertiary ($m_3$) on a wider orbit with $a_\OUT$ and $e_\OUT$.
The angular momenta of the inner and outer binaries are
denoted by $\textbf{L}_\IN\equiv\mathrm{L}_\IN\hat{\textbf{L}}_\IN$ and
$\textbf{L}_\OUT\equiv\mathrm{L}_\OUT\hat{\textbf{L}}_\OUT$ (where $\hat{\textbf{L}}_\IN$ and $\hat{\textbf{L}}_\OUT$ are unit vectors).

Gravitational perturbation from the SMBH make the inner binary precess, and may also induce Lidov-Kozai (LK) eccentricity oscillations
if the mutual inclination between $\hat{\textbf{L}}_\IN$ and
$\hat{\textbf{L}}_\OUT$ is sufficiently high.
The relevant timescale is
%%%%%%%%%%%%%%%%%%%%%%%%%%%%%%%%%%%%%%%%%%%%%%%%%%%%%%%%%%%%%%%%%%%%%%
\be\label{eq: LK timescale}
t_\lk=\frac{1}{\Omega_\lk}=\frac{1}{n_\IN}\frac{m_{12}}{m_3}\bigg(\frac{a_\OUT\sqrt{1-e^2_\OUT}}{a_\IN}\bigg)^3,
\ee
%%%%%%%%%%%%%%%%%%%%%%%%%%%%%%%%%%%%%%%%%%%%%%%%%%%%%%%%%%%%%%%%%%%%%%
where $m_{12}\equiv m_1+m_2$ and $n_\IN=(G m_{12}/a_\IN^3)^{1/2}$ is the mean motion of the inner binary.

The first-order post-Newtonian (PN) theory introduces pericenter precession in both inner and outer binaries.
In particular, the precession of the inner orbit competes with $\Omega_\lk$, and plays a crucial role in determining the
maximum eccentricity $e_\m$ in LK oscillations \citep[e.g.,][]{Fabrycky 2007,Liu et al 2015}.

Since the tertiary mass $m_3$ is much larger than the masses of the inner binary, $m_3\gg m_1, m_2$,
several general relativity (GR) effects involving the SMBH can generate extra precessions
on the binary orbits, and qualitatively change the dynamics \citep[e.g.,][]{Naoz GR,C. M. Will PRD,Liu 2019 ApJL,Liu 2020 PRD}.
In a systematical post-Newtonian framework of triple dynamics
\citep[e.g.,][]{C. M. Will PRD,Fang Yun 2019,Rodriguez 2020}, there are numerous terms.
We summarize the most essential effects below (also the leading-order effects).
The related equations are either from the classical work on binaries with spinning bodies \citep[e.g.,][]{Barker 1975},
or can be derived (or extended to include eccentricity) ``by analogy",
i.e., by recognizing that the inner binary's orbital angular momentum
$\textbf{L}_\IN$ behaves like a ``spin"\citep[][]{Liu 2019 ApJL,Liu 2020 PRD}.
As we see below, the vector forms of the equations we use are much more transparent than the equations based on orbital elements
\citep[see][]{C. M. Will PRD,Rodriguez 2020}, especially when dealing with misaligned $\textbf{L}_\IN$, $\textbf{L}_\OUT$ and $\textbf{S}_3$.
Here, we consider the double-averaged (DA; averaging over both the inner and outer orbital periods)
approximation, and present the secular equations of $\textbf{L}_\IN$ in vector forms.
The coupled eccentricity equations for various GR effects
are summarized in Appendix \ref{Appendix A}.

(i)\textit{Effect I: Precession of $\textbf{L}_\OUT$ around $\textbf{S}_3$}.
If the SMBH is rotating (with the spin angular momentum $\mathrm S_3=\chi_3G m_3^2/c$,
where $\chi_3\leqslant1$ is the Kerr parameter),
$\textbf{L}_\OUT$ experiences precession around $\textbf{S}_3$
due to spin-orbit coupling if the two vectors
are misaligned (1.5 PN effect) \citep[e.g.,][]{Barker 1975,Fang Yun}
%%%%%%%%%%%%%%%%%%%%%%%%%%%%%%%%%%%%%%%%%%%%%%%%%%%%%%%%%%%%%%%%%%%%%%
\be\label{eq:LOUT S3}
\frac{d\textbf{L}_\OUT}{dt}\bigg|_\mathrm{L_\OUT S_3}=\Omega_{\mathrm{L_\OUT S_3}}\hat{\textbf{S}}_3\times\textbf{L}_\OUT,
\ee
%%%%%%%%%%%%%%%%%%%%%%%%%%%%%%%%%%%%%%%%%%%%%%%%%%%%%%%%%%%%%%%%%%%%%%
with
%%%%%%%%%%%%%%%%%%%%%%%%%%%%%%%%%%%%%%%%%%%%%%%%%%%%%%%%%%%%%%%%%%%%%%
\be\label{eq:LOUT S3 rate}
\Omega_\mathrm{L_\OUT S_3}=\frac{GS_3(4+3m_{12}/m_3)}{2c^2a_\OUT^3(1-e_\OUT^2)^{3/2}}.
\ee
%%%%%%%%%%%%%%%%%%%%%%%%%%%%%%%%%%%%%%%%%%%%%%%%%%%%%%%%%%%%%%%%%%%%%%
For the binary-SMBH system, $\mathrm S_3$ can be easily much larger than $\mathrm L_\OUT$, and we can assume $\hat{\textbf{S}}_3=\mathrm{constant}$.

(ii)\textit{Effect II: Precession of $\textbf{L}_\IN$ around $\textbf{L}_\OUT$}.
In addition to the Newtonian precession (driven by the tidal potential of $m_3$),
$\textbf{L}_\IN$ experiences an additional de-Sitter like (geodesic) precession
in the gravitational field of $m_3$ introduced by GR (1.5 PN effect),
such that the net precession is governed by
%%%%%%%%%%%%%%%%%%%%%%%%%%%%%%%%%%%%%%%%%%%%%%%%%%%%%%%%%%%%%%%%%%%%%%
\be\label{eq:LinLout L}
\frac{d \textbf{L}_\IN}{dt}\bigg|_\mathrm{L_\IN L_\OUT}=\Omega_\mathrm{L_\IN L_\OUT}\hat{\textbf{L}}_\OUT\times\textbf{L}_\IN,
\ee
%%%%%%%%%%%%%%%%%%%%%%%%%%%%%%%%%%%%%%%%%%%%%%%%%%%%%%%%%%%%%%%%%%%%%%
with $\Omega_\mathrm{L_\IN L_\OUT}\equiv-\Omega_\mathrm{L_\IN L_\OUT}^{(\mathrm{N})}
\big(\hat{\textbf{L}}_\OUT\cdot\hat{\textbf{L}}_\IN\big)+\Omega_\mathrm{L_\IN L_\OUT}^{(\gr)}$, and
%%%%%%%%%%%%%%%%%%%%%%%%%%%%%%%%%%%%%%%%%%%%%%%%%%%%%%%%%%%%%%%%%%%%%%
\be\label{eq:GR LinLout rate}
\Omega_\mathrm{L_\IN L_\OUT}^{(\mathrm{N})}=\frac{3}{4}\Omega_\lk,~
\Omega_\mathrm{L_\IN L_\OUT}^{(\gr)}=\frac{3}{2}\frac{G (m_3+\mu_\OUT/3)n_\OUT}{c^2a_\OUT(1-e_\OUT^2)},
\ee
%%%%%%%%%%%%%%%%%%%%%%%%%%%%%%%%%%%%%%%%%%%%%%%%%%%%%%%%%%%%%%%%%%%%%%
where $n_\OUT=(Gm_\tot/a_\OUT^3)^{1/2}$.
Note that the Newtonian contribution to the precession rate neglects octuple and high-order terms;
this is justified since dynamical stability of the triple requires $a_\OUT\gg a_\IN$ when $m_3\gg m_{12}$.
For the GR part, high-order contributions to $\Omega_\mathrm{L_\IN L_\OUT}^{(\gr)}$ can be
found in \citep[e.g.,][]{C. M. Will PRD,Rodriguez 2020}.

(iii)\textit{ Effect III: Precession of $\textbf{L}_\IN$ around $\textbf{S}_3$}.
Since the semimajor axis of the inner orbit ($a_\IN$) is much smaller than the outer orbit ($a_\OUT$),
the inner binary can be treated as a single body approximately.
Thus, the angular momentum $\textbf{L}_\IN$ is coupled to the spin angular momentum $\textbf{S}_3$ of $m_3$,
and experiences Lens-Thirring precession (2 PN effect).
%%%%%%%%%%%%%%%%%%%%%%%%%%%%%%%%%%%%%%%%%%%%%%%%%%%%%%%%%%%%%%%%%%%%%%
\be\label{eq:LIN S3}
\frac{d\textbf{L}_\IN}{dt}\bigg|_\mathrm{L_\IN S_3}=\Omega_{\mathrm{L_\IN S_3}}\hat{\textbf{S}}_3\times\textbf{L}_\IN
-3\Omega_{\mathrm{L_\IN S_3}}(\hat {\textbf{L}}_\OUT\cdot \hat {\textbf{S}}_3)\hat {\textbf{L}}_\OUT\times\textbf{L}_\IN,
\ee
%%%%%%%%%%%%%%%%%%%%%%%%%%%%%%%%%%%%%%%%%%%%%%%%%%%%%%%%%%%%%%%%%%%%%%
where
%%%%%%%%%%%%%%%%%%%%%%%%%%%%%%%%%%%%%%%%%%%%%%%%%%%%%%%%%%%%%%%%%%%%%%
\be\label{eq:LIN S3 rate}
\Omega_\mathrm{L_\IN S_3}=\frac{GS_3}{2c^2a_\OUT^3(1-e_\OUT^2)^{3/2}}.
\ee
%%%%%%%%%%%%%%%%%%%%%%%%%%%%%%%%%%%%%%%%%%%%%%%%%%%%%%%%%%%%%%%%%%%%%%

Note that since the tertiary companion studied here is a SMBH,
the binary embedded in the nuclear star cluster might also be influenced by the ``environmental" effects,
including binary evaporation \citep[e.g.,][]{Tremaine}, resonant relaxation \citep[e.g.,][]{Hamers},
and non-spherical mass distribution \citep[e.g.,][]{Petrovich}.
All these effects operate on timescales much longer than considered here and can be safely neglected.

\section{Analytic Understanding of $\textbf{L}_\IN$ Evolution for Circular Inner Binary}
In \citet{Liu 2019 ApJL}, we have examined how various GR effects
modify the LK eccentricity evolution/growth of the inner binary and
enhance the merger rate in the LIGO band.
In \citet{Liu 2020 PRD}, we have studied eccentricity growth due to apsidal precession resonance in nearly co-planar triple systems.
Here, we focus on the dynamics of the inner binary far from merger, which radiates GW at low frequency band instead.
Note that the evolution of $\textbf{L}_\IN$ is more sensitive to the SMBH spin than the orbital eccentricity.
This is because the eccentricity excitation can be completely suppressed
if the mutual inclination angle lies outside the LK window, or the binary is relatively far away from
the SMBH.

\begin{figure}
\begin{centering}
\includegraphics[width=4.8cm]{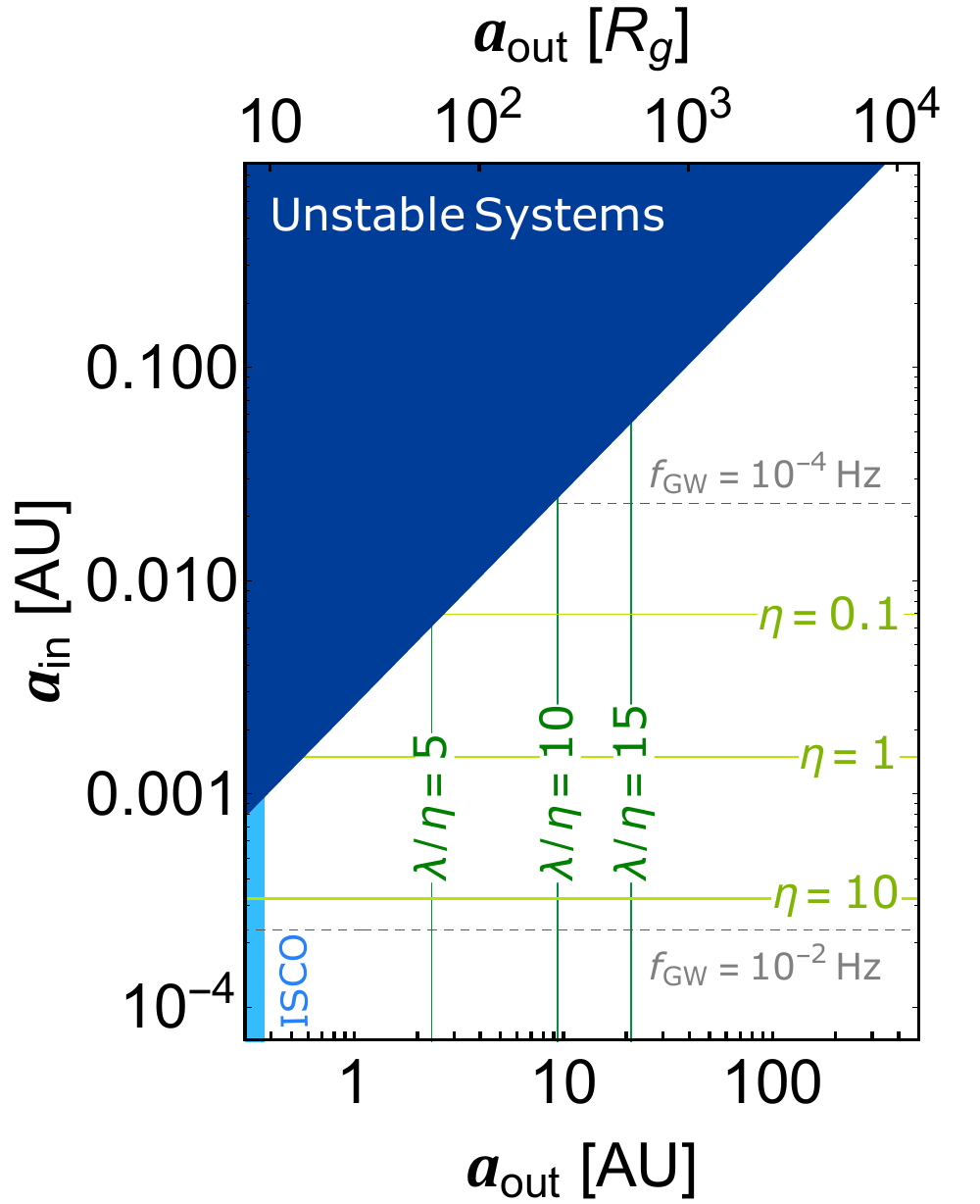}
\includegraphics[width=3.65cm]{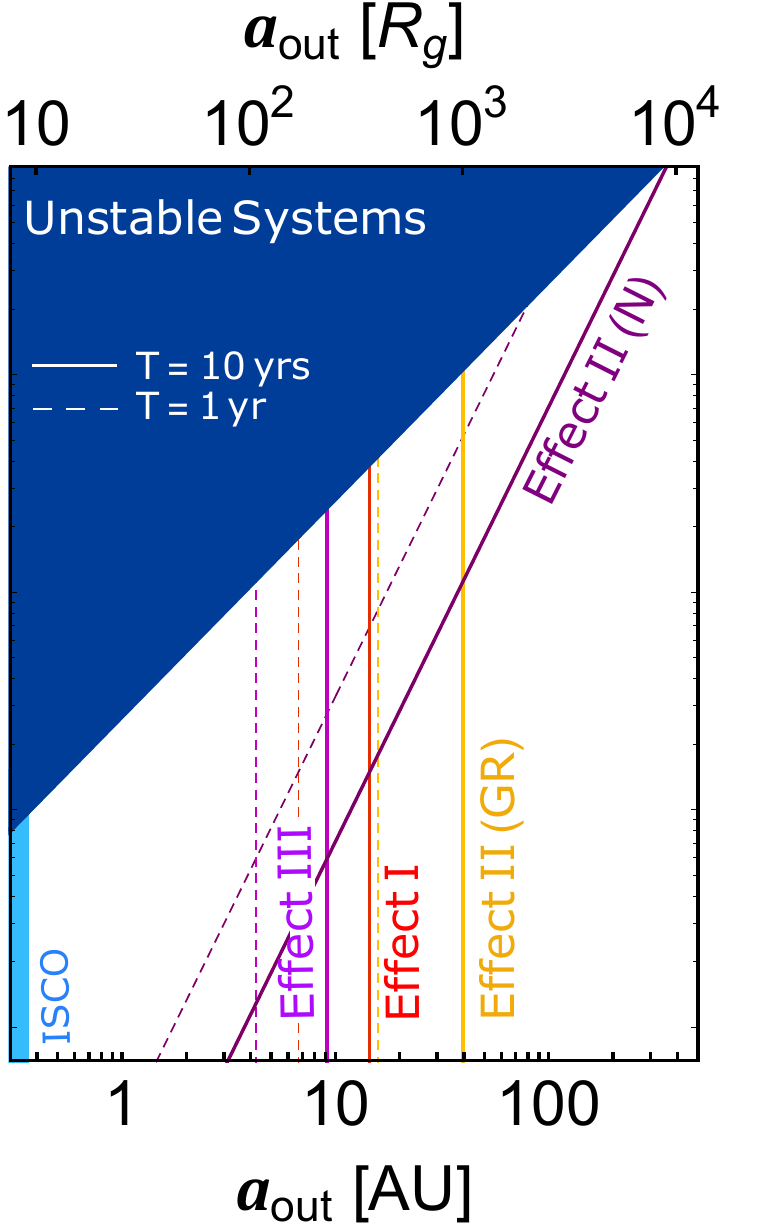}
\caption{
Parameter space in $a_\IN-a_\OUT$ plane indicating the relative importance of various GR effects.
The system parameters are $m_1=20M_\odot$, $m_2=10M_\odot$, $m_3=4\times10^6M_\odot$, $e_\IN=0$, $e_\OUT=0.5$ and $\chi_3=1$.
The dark blue region corresponds to dynamically unstable triple systems
(the instability limit according to \citet{Kiseleva 1996}), and the light blue region indicates the innermost stable circular orbits (ISCO)
for the outer binary, with $a_\OUT\leq9R_g=9(G m_3)/c^2$.
In the left panel, the solid lines show different values of $\lambda/\eta$ and $\eta$ (evaluated using Eq. \ref{eq: lambda eta}),
and the dashed lines characterize the frequency of GWs emitted by the inner binary.
In the right panel, the solid (dashed) lines are obtained
by setting the relevant timescales (Eqs. \ref{eq: LK timescale}, \ref{eq:LOUT S3 rate}, \ref{eq:GR LinLout rate} and \ref{eq:LIN S3 rate})
to 10 yrs (1 yr).
}
\label{fig:parameter space}
\end{centering}
\end{figure}

\begin{figure*}
\centering
\includegraphics[width=5.9cm]{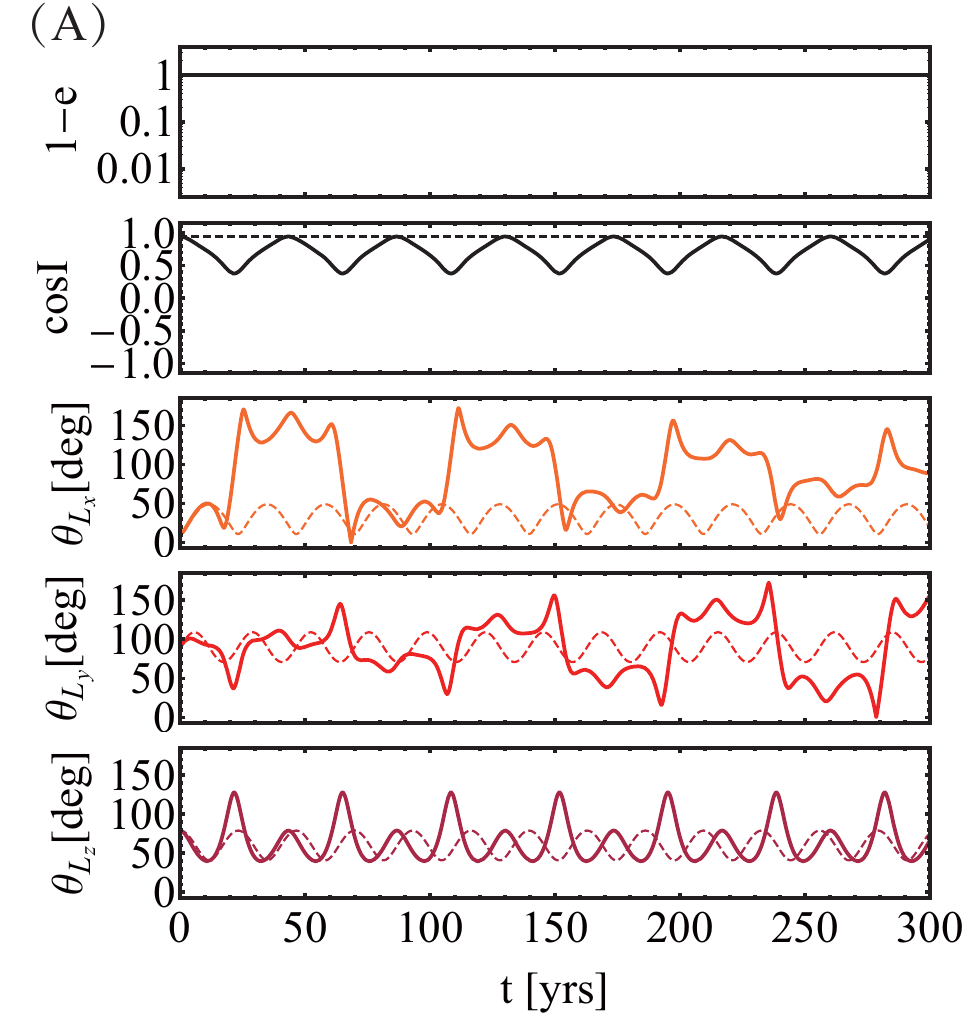}
\includegraphics[width=5.9cm]{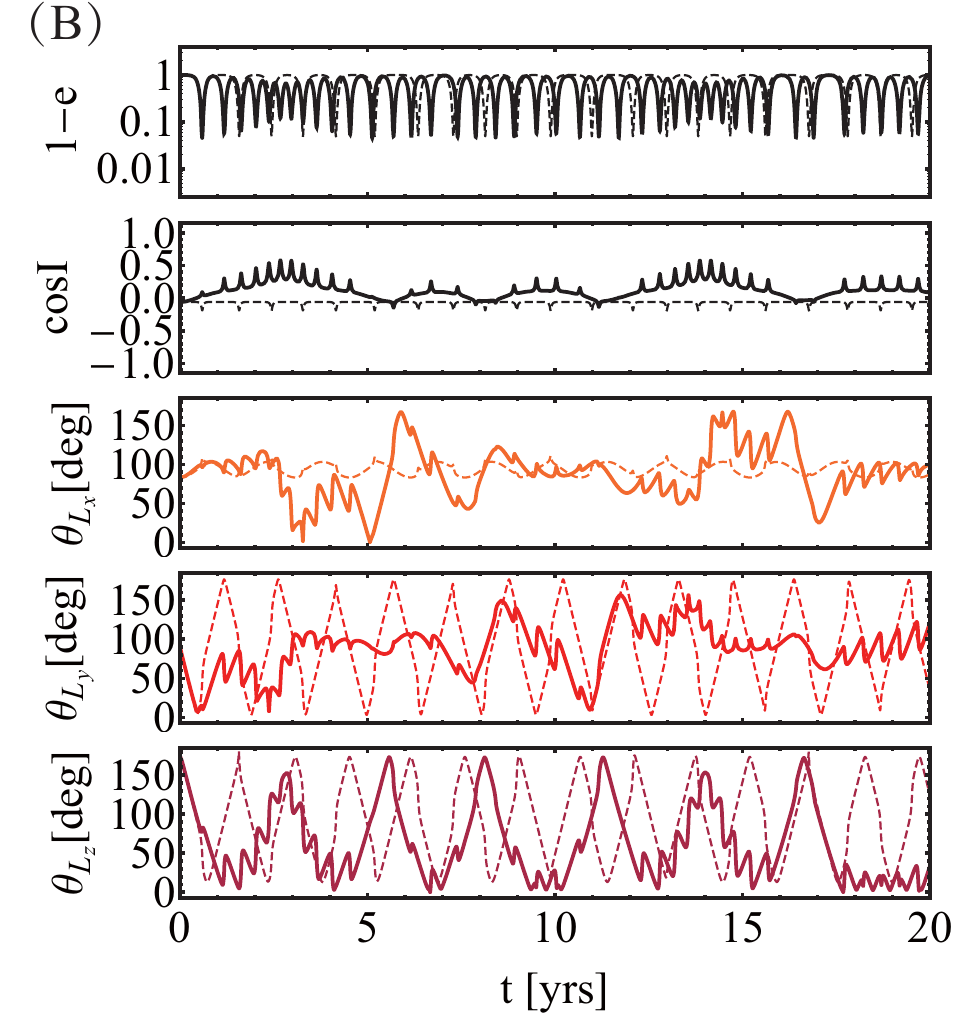}
\includegraphics[width=5.9cm]{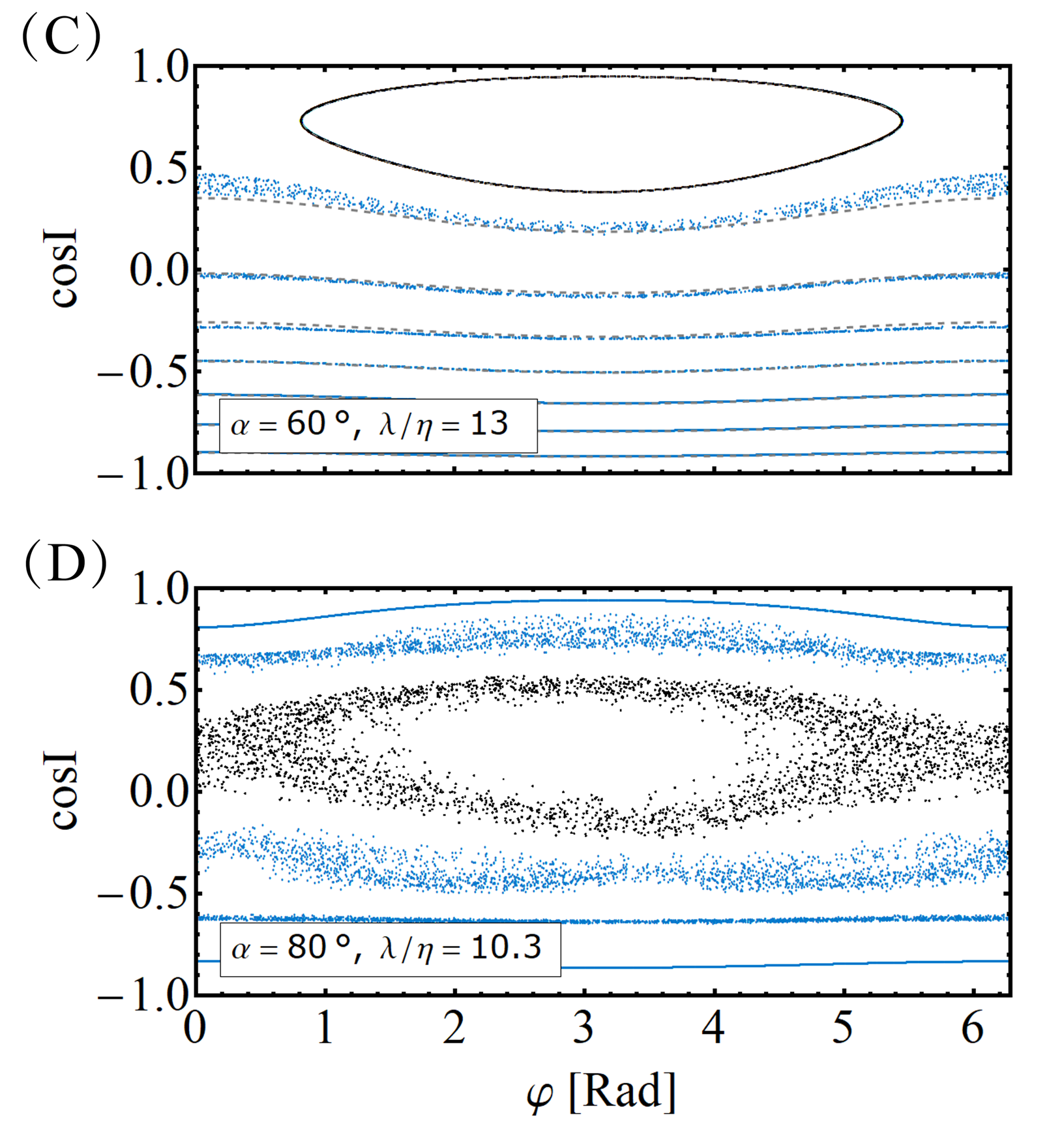}
\caption{
Sample eccentricity and angular momentum axis evolution of a BHB around a spinning SMBH tertiary.
Panels (A) and (B) show the eccentricity, inclination (the angle between ${\hat{\bf L}}_\IN$ and ${\hat{\bf L}}_\OUT$), and
the misalignment angles between $\hat {\textbf{L}}_\IN$ and the fixed x, y, z axes
(where the z-axis is aligned with $\hat {\textbf{S}}_3$ and the x-axis is in the initial $\hat {\textbf{S}}_3-\hat {\textbf{L}}_\OUT$ plane).
The parameters are $m_1=20M_\odot$, $m_2=10M_\odot$, $a_\IN=0.01\au$, $m_3=4\times10^6M_\odot$, $e_{\IN,0}=0.001$, $e_\OUT=0.5$,
$a_\OUT=15.8\au$, $I_0=19^\circ$, $\alpha=60^\circ$ (panel A) and $a_\OUT=10\au$, $I_0=93.3^\circ$, $\alpha=80^\circ$ (panel B).
The solid/dashed trajectories represent the evolution with/without the effects included by the SMBH spin (Effects II-III).
Panels (C) and (D) show the phase-space portraits obtained from the numerical integration
($\varphi$ is the precessional phase of ${\hat{\bf L}}_\IN$ around ${\hat{\bf L}}_\OUT$),
where the black dots correspond to the examples in panels (A) and (B) and the blue dots are obtained
for different values of initial inclinations (Panel (C): $I_0=79^\circ, 97^\circ, 109^\circ, 120^\circ, 131^\circ, 142^\circ, 156^\circ$;
Panel (D): $I_0=20^\circ, 30^\circ, 102^\circ, 120^\circ, 150^\circ$).
The dashed lines in panel (C) are contours of constant $\bar{\mathcal{H}}$ (Eq. \ref{eq: Hamiltonian}) for a circular inner orbit
(with $\lambda=0.74$ and $\eta=0.057$).
}
\label{fig:Orbital Evolution}
\end{figure*}

To develop an analytic understanding of the dynamics of the binary-SMBH system, we first consider the
case where the inner binary remains circular throughout the evolution.
Since $\hat {\textbf{L}}_\OUT$ rotates around $\hat{\textbf{S}}_3$ at a constant rate, $\Omega_\mathrm{L_\OUT S_3}$,
it is useful to consider the evolution of $\hat {\textbf{L}}_\IN$ in the frame corotating with $\hat {\textbf{L}}_\OUT$;
Combining Eqs. (\ref{eq:LinLout L}), (\ref{eq:LOUT S3}) and (\ref{eq:LIN S3}), we have
%%%%%%%%%%%%%%%%%%%%%%%%%%%%%%%%%%%%%%%%%%%%%%%%%%%%%%%%%%%%%%%%%%%%%%
\ba
\bigg(\frac{d\textbf{L}_\IN}{dt}\bigg)_\mathrm{rot}=&&\bigg\{
\Big[\Omega_\mathrm{L_\IN L_\OUT}-3\Omega_{\mathrm{L_\IN S_3}}(\hat {\textbf{L}}_\OUT\cdot \hat {\textbf{S}}_3)\Big]\hat {\textbf{L}}_\OUT\nonumber\\
&&+\Big(\Omega_{\mathrm{L_\IN S_3}}-\Omega_{\mathrm{L_\OUT S_3}}\Big)\hat{\textbf{S}}_3\bigg\}\times\textbf{L}_\IN.
\ea
%%%%%%%%%%%%%%%%%%%%%%%%%%%%%%%%%%%%%%%%%%%%%%%%%%%%%%%%%%%%%%%%%%%%%%
The corresponding Hamiltonian is
%%%%%%%%%%%%%%%%%%%%%%%%%%%%%%%%%%%%%%%%%%%%%%%%%%%%%%%%%%%%%%%%%%%%%%
\ba
\mathcal{H}=&&-\frac{1}{2}\Omega_\mathrm{L_\IN L_\OUT}^{(\mathrm{N})}
(\hat {\textbf{L}}_\IN\cdot\hat {\textbf{L}}_\OUT)^2+\Omega_\mathrm{L_\IN L_\OUT}^{(\mathrm{GR})}(\hat {\textbf{L}}_\IN\cdot\hat {\textbf{L}}_\OUT)\nonumber\\
&&+(\Omega_{\mathrm{L_\IN S_3}}-\Omega_{\mathrm{L_\OUT S_3}})
(\hat {\textbf{L}}_\IN\cdot\hat {\textbf{S}}_3)\nonumber\\
&&-3\Omega_{\mathrm{L_\IN S_3}}
(\hat {\textbf{L}}_\OUT\cdot\hat {\textbf{S}}_3)(\hat {\textbf{L}}_\IN\cdot\hat {\textbf{L}}_\OUT).
\ea
%%%%%%%%%%%%%%%%%%%%%%%%%%%%%%%%%%%%%%%%%%%%%%%%%%%%%%%%%%%%%%%%%%%%%%
We set up a coordinate system with $\hat z=\hat {\textbf{L}}_\OUT$,
$\hat y\sin\alpha\equiv\hat {\textbf{L}}_\OUT\times\hat {\textbf{S}}_3$, and let
$\hat {\textbf{L}}_\IN=\sin I(\cos\varphi\hat x+\sin\varphi\hat y)+\cos I\hat z$, where
$\alpha$ is the angle between $\hat {\textbf{L}}_\OUT$ and $\hat {\textbf{S}}_3$, and
$I$ is the angle between $\hat {\textbf{L}}_\IN$ and $\hat {\textbf{L}}_\OUT$. The (dimensionless) Hamiltonian becomes
%%%%%%%%%%%%%%%%%%%%%%%%%%%%%%%%%%%%%%%%%%%%%%%%%%%%%%%%%%%%%%%%%%%%%
\ba\label{eq: Hamiltonian}
\bar{\mathcal{H}}=\frac{\mathcal{H}}{\Omega_\mathrm{L_\IN L_\OUT}^{(\mathrm{N})}}
=&&-\frac{1}{2}\cos^2I+\lambda\cos I\\
&&-\frac{3}{4}\eta\big(2\cos\alpha\cos I+\sin\alpha\sin I\cos\varphi\big),\nonumber
\ea
%%%%%%%%%%%%%%%%%%%%%%%%%%%%%%%%%%%%%%%%%%%%%%%%%%%%%%%%%%%%%%%%%%%%%%
where we have introduced the dimensionless ratios
%%%%%%%%%%%%%%%%%%%%%%%%%%%%%%%%%%%%%%%%%%%%%%%%%%%%%%%%%%%%%%%%%%%%%%
\be\label{eq: lambda eta}
\lambda=\frac{\Omega_\mathrm{L_\IN L_\OUT}^{(\mathrm{GR})}}{\Omega_\mathrm{L_\IN L_\OUT}^{(\mathrm{N})}},~~~
\eta=\frac{\Omega_\mathrm{L_\OUT S_3}}{\Omega_\mathrm{L_\IN L_\OUT}^{(\mathrm{N})}},
\ee
%%%%%%%%%%%%%%%%%%%%%%%%%%%%%%%%%%%%%%%%%%%%%%%%%%%%%%%%%%%%%%%%%%%%%%
and have used $\Omega_\mathrm{L_\IN S_3}/\Omega_\mathrm{L_\OUT S_3}=(4+3m_{12}/m_3)^{-1}\simeq1/4$.
Note that $\cos I$ and $\varphi$ are canonical variables.

Depending on the ratio of $\lambda$ and $\eta$, we expect three possible $\hat {\textbf{L}}_\IN$ behaviors
\footnote{Here we introduce $\lambda$ and $\eta$ to quantify the $\hat {\textbf{L}}_\IN$ behavior
and the resonance. Note that we have defined a similar factor $\gamma$ in \cite{Liu 2019 ApJL},
which involves the dependence on the inclination ($I$)}:
(i) For $|\lambda\pm1|\gg\eta$ (``adiabatic"), $\hat {\textbf{L}}_\IN$ closely follows $\hat {\textbf{L}}_\OUT$,
maintaining an approximately constant $I$;
(ii) For $|\lambda\pm1|\ll\eta$ (``nonadiabatic"), $\hat {\textbf{L}}_\IN$ effectively precesses around $\hat {\textbf{S}}_3$
with constant $\theta_\mathrm{L_\IN, S_3}$ (the angle between $\hat {\textbf{L}}_\IN$ and $\hat {\textbf{S}}_3$);
(iii) When $|\lambda\pm1|\sim\eta$ (``trans-adiabatic"),
a resonance behavior of $\hat {\textbf{L}}_\IN$ may occur,
and large orbital inclination $I$ can be generated.

Fig. \ref{fig:parameter space} presents the parameter space indicating the relative importance of various GR effects
for compact BH binaries (BHBs) around SgrA$^\ast$.
In the left panel, we see that
for BHBs ($m_1=20M_\odot$ and $m_2=10M_\odot$) that radiate GWs in the low-frequency band ($10^{-4}-10^{-2}$Hz),
$\eta$ ranges from $0.02$ to 10. The ``nonadiabatic" parameter regime ($\lambda\ll\eta$) is not allowed
for the realistic systems because of the stability criterion and the effect of ISCO (Inner-most stable circular orbit).
As $a_\OUT$ increases, $\lambda/\eta$ increases and the dynamics of $\hat {\textbf{L}}_\IN$ transitions from
``trans-adiabatic" to ``adiabatic". Thus,
resonance behavior of $\hat {\textbf{L}}_\IN$ (i.e., $|\lambda\pm1|\sim\eta$)
can only occur when $\eta\lesssim1$.
In addition, to obtain variations of the orbit on relatively short timescales ($\lesssim10$yrs;
see the right panel of Fig. \ref{fig:parameter space}),
the BHB cannot be too far away from the SMBH (i.e., $a_\OUT\lesssim50$AU).

Fig. \ref{fig:Orbital Evolution} (panel A) illustrates how the GR effects induced by the spinning SMBH modify the evolution of
$\hat {\textbf{L}}_\IN$ of a BHB.
We find that the orbital inclination $I$ undergoes significant change due to the spin effects (Effects II-III; $\chi_3=1$),
and the misalignment angles $\theta_{\mathrm{L}_\mathrm{x,y,z}}$ between $\hat {\textbf{L}}_\IN$ and the fixed x, y, z axes
exhibit dramatic oscillations. For reference,
for a non-spinning SMBH, $I$ stays constant and only regular oscillations of $\theta_{\mathrm{L}_\mathrm{x,y,z}}$
with small amplitude are produced.

Panel (C) of Fig. \ref{fig:Orbital Evolution} shows the evolution of $\hat {\textbf{L}}_\IN$ in the
($\varphi$, $\cos I$) phase space. We see that the large variation of inclination is associated with the
librating trajectory (i.e., resonance phenomenon),
which is well described by the Hamiltonian (Eq. \ref{eq: Hamiltonian}) with $\lambda=0.74$ and $\eta=0.057$.

\begin{figure}
\includegraphics[width=8.5cm]{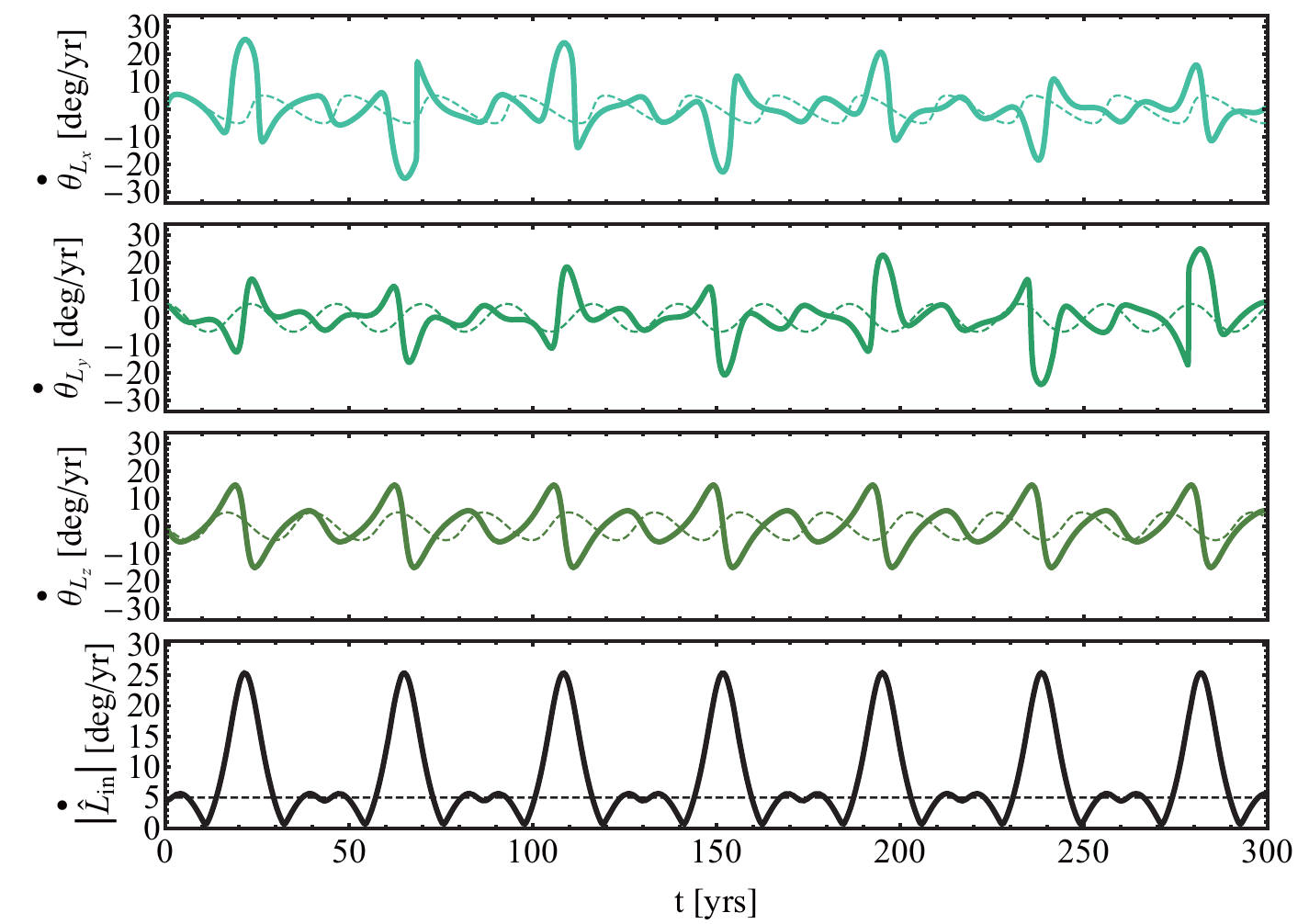}
\caption{The rate of change of $\hat {\textbf{L}}_\IN$ and the misalignment angles $\theta_{\mathrm{L}_\mathrm{x,y,z}}$
(the angle between $\hat {\textbf{L}}_\IN$ and the fixed x, y, z axes) for the BHB evolution
depicted in Fig. \ref{fig:Orbital Evolution} (A). Here, the z-axis is along $\hat {\textbf{S}}_3$,
and the y-axis is along the initial ($\hat {\textbf{S}}_3\times\hat {\textbf{L}}_\OUT$) direction.
}
\label{fig:Rate of  Change}
\end{figure}

As the BHB precesses, the GW waveform undergoes both amplitude
and phase modulations, thereby allowing the measurement of the change in the orientation of $\hat {\textbf{L}}_\IN$ \citep[e.g.,][]{Yuhang PRL}.
Since $\theta_{\mathrm{L}_\mathrm{x,y,z}}$ evolves irregularly on the secular timescale that
could be longer than the mission lifetime of future GW detectors (such as LISA),
we recognize that the rate of change of $\hat {\textbf{L}}_\IN$ (and of $\theta_{\mathrm{L}_\mathrm{x,y,z}}$)
would be a useful observable indicator to track the evolution of the BHB.
Fig. \ref{fig:Rate of Change} shows the evolution of $d\theta_{\mathrm{L}_\mathrm{x,y,z}}/dt$ and $d\hat {\textbf{L}}_\IN/dt$
for the example depicted in Fig. \ref{fig:Orbital Evolution} (A).
We see that the overall oscillations of $d\theta_{\mathrm{L}_\mathrm{x,y,z}}/dt$ and $d\hat {\textbf{L}}_\IN/dt$ induced by the SMBH spin effects are more
dramatic than the case without the spin effects,
with $d\theta_{\mathrm{L}_\mathrm{x}}/dt$ and $d\hat {\textbf{L}}_\IN/dt$ reaching an amplitude of $25^\circ/$yr
--- Such a large and rapid change in $\hat {\textbf{L}}_\IN$ leads to
significant amplitude modulation of the gravitational waveform, and should be easily detectable.
In contrast, without the spin effects of the SMBH,
\be\label{eq:dot L DS}
\bigg|\frac{d\hat {\textbf{L}}_\IN}{dt}\bigg|_{\chi_3=0;e_\IN=0}=
\big|\Omega_\mathrm{L_\IN L_\OUT}^{(\mathrm{GR})}-\Omega_\mathrm{L_\IN L_\OUT}^{(\mathrm{N})}\cos I_0\big|\sin I_0,
\ee
is a constant ($\simeq5^\circ/$yr; see the bottom panel of Fig. \ref{fig:Rate of  Change}).

\section{General Case}
The BHB may experience eccentricity growth through LK oscillations when the inclination $I$ is sufficiently large.
The precession of $\hat {\textbf{L}}_\OUT$ around $\hat {\textbf{S}}_3$ can increase the inclination window of
eccentricity excitation \citep[e.g.,][]{Liu 2019 ApJL}.
In this situation, the finite eccentricity of the inner binary increases the GW strain, which affects the overall signal-to-noise ratio
and improves the detectability \citep[e.g.,][]{Xianyu LISA,Hoang,Deme}.

\begin{figure}
\includegraphics[width=9cm]{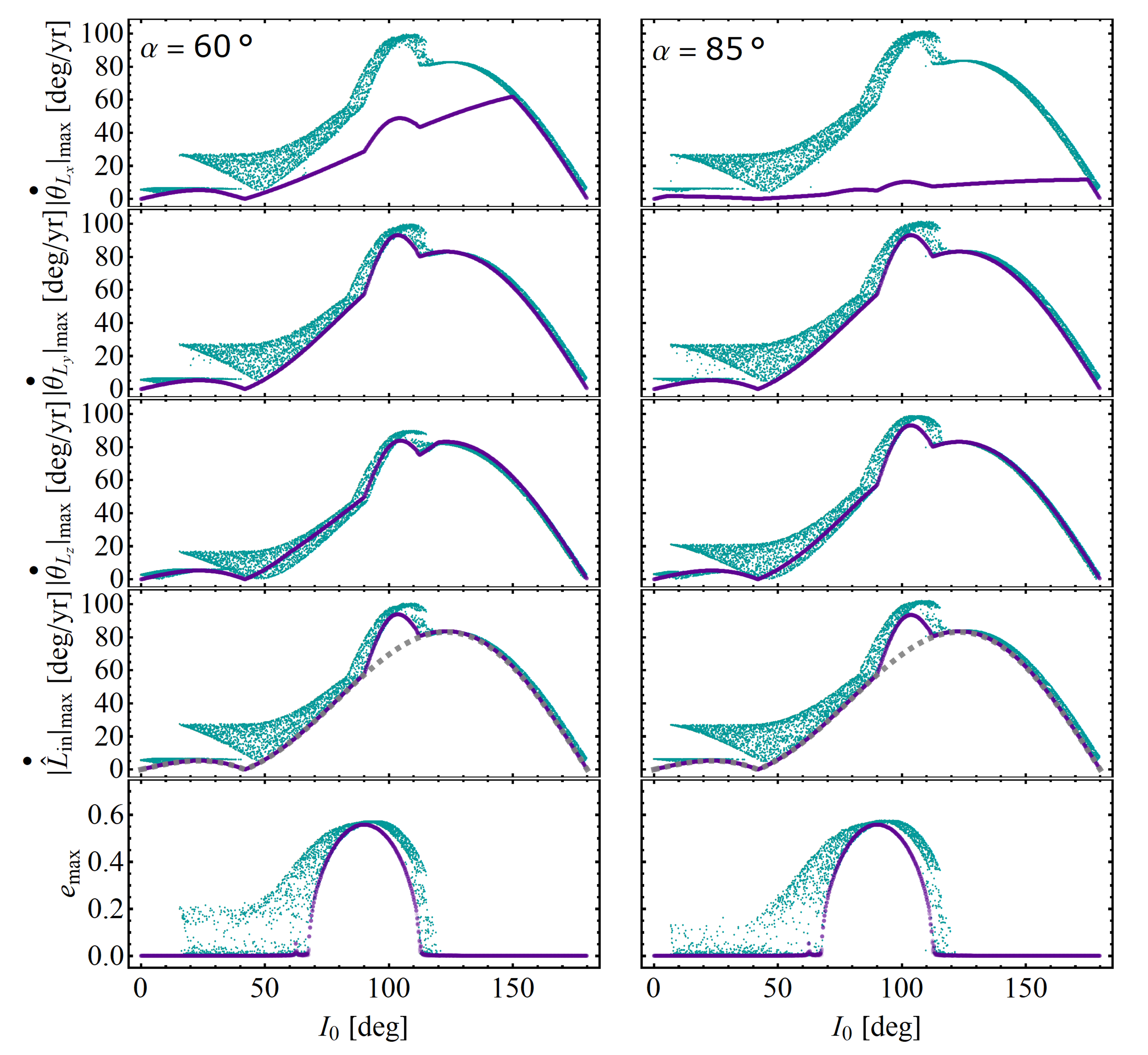}
\caption{
Maximum rates of change of ${\theta}_{\mathrm{L}_{x,y,z}}$, ${\hat {\textbf{L}}}_\IN$ and
the maximum eccentricity of the inner BHB vs. the initial inclination $I_0$
for two different misalignment angle between $\hat {\textbf{S}}_3$ and $\hat {\textbf{L}}_\OUT$ ($\alpha$; as labeled).
The inner binary has $m_1=20M_\odot$, $m_2=10M_\odot$, $a_\IN=0.01$AU, and the SMBH has
$m_3=4\times10^6M_\odot$, $e_\OUT=0.5$ and $a_\OUT=15.8\au$ (the initial eccentricity $e_\IN=0.001$).
Each $I_0$ is simulated ten times, with the initial orbital elements (the arguments of
pericenters, the longitudes of the ascending nodes and the azimuthal phase angle of $\hat {\textbf{S}}_3$)
chosen randomly from 0 to $2\pi$ for each integration.
We evolve the DA secular equations for the triple systems for 2000 yrs.
The purple (cyan) dots are the results without (with) the SMBH spin effects.
The dashed lines in the panels of $|\dot{\hat {\textbf{L}}}_\IN|_\m$
are from Eq. (\ref{eq:dot L DS})
assuming circular inner binary and non-spinning SMBH.
}
\label{fig:Spin window}
\end{figure}

The left panels of Fig. \ref{fig:Spin window} show the maximum values of $d\theta_{\mathrm{L}_\mathrm{x,y,z}}/dt$, $|d\hat {\textbf{L}}_\IN/dt|$
and the maximum eccentricity $e_\m$ as a function of the initial inclination $I_0$ for the same system parameters as
in Fig. \ref{fig:Orbital Evolution} (A).
We see that in the absent of spin effects (purple dots), the maximum rates and eccentricity
are uniquely determined by $I_0$, and
$|d\hat {\textbf{L}}_\IN/dt|$ agrees with Eq. (\ref{eq:dot L DS}) for a wide range of $I_0$
even when the inner binary develops eccentricities
(this arises because $|d\hat {\textbf{L}}_\IN/dt|_\m$ is achieved when $e_\IN\simeq0$ during the LK cycles).
However, with the inclusion of the SMBH spin effects (cyan dots), the eccentricity excitation window is widen \citep[e.g.,][]{Liu 2019 ApJL},
and there can be a finite spread of the maximum values of $|d\theta_{\mathrm{L}_\mathrm{x,y,z}}/dt|$ and $|d\hat {\textbf{L}}_\IN/dt|$ for each $I_0$.
Note that for systems with $e_\m\lesssim0.6$,
the dynamics of $\hat {\textbf{L}}_\IN$
can still be described approximately in an analytical way, using the ``circular" Hamiltonian (Eq. \ref{eq: Hamiltonian}).
This has been seem in Fig. \ref{fig:Orbital Evolution} (C):
the numerical trajectories (blue dots) are close to the analytical $\bar{\mathcal{H}}=\mathrm{constant}$ curves.

To explore the dependence of the direction of $\hat {\textbf{S}}_3$ on the evolution of $\hat {\textbf{L}}_\IN$,
we consider a different $\alpha$ value ($=85^\circ$), and the
results are shown in the right panels of Fig. \ref{fig:Spin window}.
Similar distributions are obtained, except that the $e-$excitation window is broader
than the $\alpha=60^\circ$ case shown in the left panel
(see more details in the Supplemental Material).

The eccentricity excitation can be more significant if the BHB is closer to the SMBH.
At the same time, the precession timescales become shorter and
a wide range of variations of the BHB orbit can be potentially captured
during the observational span of a few years.
Panel (B) of Fig. \ref{fig:Orbital Evolution} shows the example with $a_\OUT=10$AU.
We see that because of the SMBH spin effects, the normal periodic oscillations in $e$ and $\cos I$ transform into irregular oscillations.
In this case, the ``circular" Hamiltonian (Eq. \ref{eq: Hamiltonian}) no longer applies.
Instead, as indicated in panel (D) of Fig. \ref{fig:Orbital Evolution}, a large degree of scatter fills up the phase space,
and the variation of $\hat {\textbf{L}}_\IN$ becomes chaotic (see Appendix \ref{Appendix B}).

\section{Summary and Discussion}
We have studied the effects of the spin of SMBH (such as SgrA$^\ast$) on the evolution of the orbital axis ($\hat {\textbf{L}}_\IN$)
of surrounding compact binaries.
We find that for typical BHBs ($m_1\sim20M_\odot$ and $m_2\sim10M_\odot$) that
are close to the SMBH ($a_\OUT\lesssim50$AU), $\hat {\textbf{L}}_\IN$ may experience complex (and even resonant)
evolution, leading to significant variation of $\hat {\textbf{L}}_\IN$.
For the BHBs that remain circular during
the evolution, this variation of $\hat {\textbf{L}}_\IN$ can be calculated analytically.
We show that a spinning SMBH can greatly influence the variation of $\hat {\textbf{L}}_\IN$
(even for the circular BHBs), increasing $|d\hat {\textbf{L}}_\IN/dt|$ significantly compared to
the case of a non-spinning SMBH.
The maximum $|d\hat {\textbf{L}}_\IN/dt|$ can reach many tens of degrees per year for BHBs
emitting GWs in the low-frequency band ($10^{-3}-10^{-1}$Hz).
Such rapid variation of $\hat {\textbf{L}}_\IN$
therefore provides a probe on the mass and spin of the SMBH.

The SMBH spin can also affect other type of compact binaries,
including neutron star binaries and white dwarf binaries.
Although there is no direct observational evidence for their existence near SgrA$^\ast$,
massive stars with distances within $\sim13$ AU from SgrA$^\ast$ are known
\citep[e.g.,][]{S star 1,S star 2,S star 3},
and it is plausible to expect stellar binaries their remnants to exist at such distances.
In addition, various dynamical processes can lead to enhanced production of compact binaries around SMBHs,
including gravitational bremsstrahlung \citep[e.g.,][]{OLeary 2009},
mass segregation \citep[e.g.,][]{Antonini 2016,Leigh 2018,Fragione 2018,Sari 2019,Arca Sedda 2020},
scatterings via eccentric disks \citep[e.g.,][]{Generozov 2020}, and tidal/GW captute \citep[e.g.,][]{Chen Xian}.

For the detectability in GWs, the types of compact binaries studied here are luminous low-frequency GW sources in the Galaxy
(e.g., the signal-to-noise ratio $\gtrsim$37 with LISA's sensitivity; see also Appendixes \ref{Appendix C} and \ref{Appendix D}).
Since the orbital period of the outer binary is much shorter than the duration of GW detection, the
system parameters can be well constrained through Doppler phase shift \citep[e.g.,][]{Inayoshi,Xianyu}.
Detecting the GW signal containing the signature of the frame-dragging effect is more challenging.
A recent study \citep[][]{Yuhang PRL} (which neglects the frame-dragging effect)
found that the regular precession of $\textbf{L}_\IN$ around $\textbf{L}_\OUT$ due to Newtonian torques can be measurable if the
precession period is less than the observation time. Similar detectability
is expected to apply for the non-regular evolution of $\hat {\textbf{L}}_\IN$ discussed in this paper.
If the system happens to be observed during the time when
$|d\hat {\textbf{L}}_\IN/dt|$ is significantly enhanced, the effect would be more ``visible".

To conclude, our
proof-of-concept calculations demonstrate that the SMBH spin can have large inprint on the BHB waveforms.
A joint detection of multiple compact binary systems
may be necessary to reduce the degeneracy of the GW signals on various parameters,
and provide sufficient constraints on the SMBH spin.
Future studies on detailed strategy to measure the SMBH spin using low-frequency GWs from
compact binaries would be of great value.

\section{Acknowledgments}

BL thank Johan Samsing and Daniel D'Orazio for useful discussion.
DL has been supported in part by NSF grants AST-1715246 and AST-2107796.
This project has received funding from the European Union's Horizon 2020
research and innovation program under the Marie Sklodowska-Curie grant agreement No. 847523 `INTERACTIONS'.

\appendix
\section{A: GR effects due to Rotating SMBH}
\label{Appendix A}

We summarize the most essential GR effects for the BHB-SMBH triple system below.
The related equations follow from the double-averaged (DA; averaging over both the inner and outer orbital periods) approximation.

(i)\textit{Effect I: Precession of $\textbf{L}_\OUT$ around $\textbf{S}_3$}.
In the BHB-SMBH system, the angular momentum of the outer binary $\textbf{L}_\OUT$
and the spin angular momentum $\textbf{S}_3$ of $m_3$
precesses around each other due to spin-orbit coupling if the two vectors
are misaligned (1.5 PN effect) \citep[e.g.,][]{Barker 1975,Fang Yun}:
%%%%%%%%%%%%%%%%%%%%%%%%%%%%%%%%%%%%%%%%%%%%%%%%%%%%%%%%%%%%%%%%%%%%%%
\ba
\frac{d\textbf{L}_\OUT}{dt}\bigg|_\mathrm{L_\OUT S_3}=&&\Omega_{\mathrm{L_\OUT S_3}}\hat{\textbf{S}}_3\times\textbf{L}_\OUT, \\
\frac{d\textbf{e}_\OUT}{dt}\bigg|_\mathrm{L_\OUT S_3}=&&\Omega_{\mathrm{L_\OUT S_3}}\hat{\textbf{S}}_3\times\textbf{e}_\OUT\nonumber\\
&&-3\Omega_{\mathrm{L_\OUT S_3}}(\hat {\textbf{L}}_\OUT\cdot \hat {\textbf{S}}_3)\hat {\textbf{L}}_\OUT\times\textbf{e}_\OUT, \\
\frac{d\hat{\textbf{S}}_3}{dt}\bigg|_\mathrm{S_3 L_\OUT}=&&\Omega_{\mathrm{S_3 L_\OUT}}\hat{\textbf{L}}_\OUT\times\hat{\textbf{S}}_3,\label{eq:S3 S3}
\ea
%%%%%%%%%%%%%%%%%%%%%%%%%%%%%%%%%%%%%%%%%%%%%%%%%%%%%%%%%%%%%%%%%%%%%%
where the orbit-averaged precession rates are
%%%%%%%%%%%%%%%%%%%%%%%%%%%%%%%%%%%%%%%%%%%%%%%%%%%%%%%%%%%%%%%%%%%%%%
\be
\Omega_\mathrm{L_\OUT S_3}=\frac{GS_3(4+3m_{12}/m_3)}{2c^2a_\OUT^3(1-e_\OUT^2)^{3/2}}=
\Omega_{\mathrm{S_3 L_\OUT}}\frac{S_3}{L_\OUT}.
\ee
%%%%%%%%%%%%%%%%%%%%%%%%%%%%%%%%%%%%%%%%%%%%%%%%%%%%%%%%%%%%%%%%%%%%%%
Since in our case, $\mathrm S_3$ can be easily larger than $\mathrm L_\OUT$, the de-Sitter precession (Eq. \ref{eq:S3 S3})
is negligible.

(ii)\textit{Effect II: Precession of $\textbf{L}_\IN$ around $\textbf{L}_\OUT$}.
In addition to the Newtonian precession (driven by the tidal potential of $m_3$),
$\textbf{L}_\IN$ experiences an additional de-Sitter like (geodesic) precession
in the gravitational field of $m_3$ introduced by GR.
This is a 1.5 PN spin-orbit coupling effect, with $\textbf{L}_\IN$ behaving like a ``spin". We have
%%%%%%%%%%%%%%%%%%%%%%%%%%%%%%%%%%%%%%%%%%%%%%%%%%%%%%%%%%%%%%%%%%%%%
\ba
&&\frac{d \textbf{L}_\IN}{dt}\bigg|_\mathrm{L_\IN L_\OUT}=\Omega_\mathrm{L_\IN L_\OUT}^{(\gr)}\hat{\textbf{L}}_\OUT\times\textbf{L}_\IN,\\
&&\frac{d \textbf{e}_\IN}{dt}\bigg|_\mathrm{L_\IN L_\OUT}=\Omega_\mathrm{L_\IN L_\OUT}^{(\gr)}\hat{\textbf{L}}_\OUT\times\textbf{e}_\IN,
\ea
%%%%%%%%%%%%%%%%%%%%%%%%%%%%%%%%%%%%%%%%%%%%%%%%%%%%%%%%%%%%%%%%%%%%%%
and the feedback from $\hat{\mathbf{L}}_\IN$, $\mathbf{e}_\IN$ on $\mathbf{L}_\OUT$ and $\mathbf{e}_\OUT$ are given by \citep[e.g.,][]{Barker 1975}
%%%%%%%%%%%%%%%%%%%%%%%%%%%%%%%%%%%%%%%%%%%%%%%%%%%%%%%%%%%%%%%%%%%%%
\ba
\frac{d \textbf{L}_\OUT}{dt}\bigg|_\mathrm{L_\OUT L_\IN}=&&\Omega_\mathrm{L_\OUT L_\IN}^{(\gr)}
\hat{\textbf{L}}_\IN\times\textbf{L}_\OUT,\\
\frac{d \textbf{e}_\OUT}{dt}\bigg|_\mathrm{L_\OUT L_\IN}=&&\Omega_\mathrm{L_\OUT L_\IN}^{(\gr)}\hat{\textbf{L}}_\IN\times\textbf{e}_\OUT\\
&&-3\omega_\mathrm{L_\OUT L_\IN}^{(\gr)}(\hat {\textbf{L}}_\OUT\cdot \hat {\textbf{L}}_\IN)\hat {\textbf{L}}_\OUT\times\textbf{e}_\OUT, \nonumber
\ea
%%%%%%%%%%%%%%%%%%%%%%%%%%%%%%%%%%%%%%%%%%%%%%%%%%%%%%%%%%%%%%%%%%%%%%
with
%%%%%%%%%%%%%%%%%%%%%%%%%%%%%%%%%%%%%%%%%%%%%%%%%%%%%%%%%%%%%%%%%%%%%%
\be
\Omega_\mathrm{L_\IN L_\OUT}^{(\gr)}=\frac{3}{2}\frac{G (m_3+\mu_\OUT/3)n_\OUT}{c^2a_\OUT(1-e_\OUT^2)}
=\Omega_\mathrm{L_\OUT L_\IN}^{(\gr)}\frac{L_\OUT}{L_\IN},
\ee
%%%%%%%%%%%%%%%%%%%%%%%%%%%%%%%%%%%%%%%%%%%%%%%%%%%%%%%%%%%%%%%%%%%%%%
where $n_\OUT=(Gm_\tot/a_\OUT^3)^{1/2}$.

(iii)\textit{ Effect III: Precession of $\textbf{L}_\IN$ around $\textbf{S}_3$}.
Since the semimajor axis of the inner orbit ($a_\IN$) is much smaller than the outer orbit ($a_\OUT$),
the inner binary can be treated as a single body approximately.
Therefore, the angular momentum $\textbf{L}_\IN$ is coupled to the spin angular momentum $\textbf{S}_3$ of $m_3$,
and experiences Lens-Thirring precession. This is a 2 PN spin-spin coupling effect, with $\textbf{L}_\IN$ behaving like a ``spin". We have
%%%%%%%%%%%%%%%%%%%%%%%%%%%%%%%%%%%%%%%%%%%%%%%%%%%%%%%%%%%%%%%%%%%%%%
\ba
\frac{d\textbf{L}_\IN}{dt}\bigg|_\mathrm{L_\IN S_3}=&&\Omega_{\mathrm{L_\IN S_3}}\hat{\textbf{S}}_3\times\textbf{L}_\IN\nonumber\\
&&-3\Omega_{\mathrm{L_\IN S_3}}(\hat {\textbf{L}}_\OUT\cdot \hat {\textbf{S}}_3)\hat {\textbf{L}}_\OUT\times\textbf{L}_\IN,\\
\frac{d\textbf{e}_\IN}{dt}\bigg|_\mathrm{L_\IN S_3}=&&\Omega_{\mathrm{L_\IN S_3}}\hat{\textbf{S}}_3\times\textbf{e}_\IN\nonumber\\
&&-3\Omega_{\mathrm{L_\IN S_3}}(\hat {\textbf{L}}_\OUT\cdot \hat {\textbf{S}}_3)\hat {\textbf{L}}_\OUT\times\textbf{e}_\IN.
\ea
%%%%%%%%%%%%%%%%%%%%%%%%%%%%%%%%%%%%%%%%%%%%%%%%%%%%%%%%%%%%%%%%%%%%%%
The back-reaction on the outer binary gives (see Eqs. 64, 65, 70 of \cite{Barker 1975})
%%%%%%%%%%%%%%%%%%%%%%%%%%%%%%%%%%%%%%%%%%%%%%%%%%%%%%%%%%%%%%%%%%%%%%
\ba
\frac{d\textbf{L}_\OUT}{dt}\bigg|_\mathrm{S_3 L_\IN}&&=-3\Omega_{\mathrm{S_3 L_\IN}}
\Big[(\hat {\textbf{L}}_\OUT\cdot\hat {\textbf{L}}_\IN)\hat {\textbf{S}}_3
+(\hat {\textbf{L}}_\OUT\cdot\hat {\textbf{S}}_3)\hat {\textbf{L}}_\IN\Big]\nonumber\\
&&\times\textbf{L}_\OUT,\\
\frac{d\textbf{e}_\OUT}{dt}\bigg|_\mathrm{S_3 L_\IN}&&=-3\Omega_{\mathrm{S_3 L_\IN}}
\bigg\{(\hat {\textbf{L}}_\OUT\cdot\hat {\textbf{L}}_\IN)\hat {\textbf{S}}_3+(\hat {\textbf{L}}_\OUT\cdot\hat {\textbf{S}}_3)
\hat {\textbf{L}}_\IN\nonumber\\
&&+\Big[(\hat {\textbf{L}}_\IN\cdot\hat {\textbf{S}}_3)-5(\hat {\textbf{L}}_\OUT\cdot\hat {\textbf{L}}_\IN)
(\hat {\textbf{L}}_\OUT\cdot\hat {\textbf{S}}_3)\Big]\hat {\textbf{L}}_\OUT\bigg\}\nonumber\\
&&\times\textbf{e}_\OUT.
\ea
%%%%%%%%%%%%%%%%%%%%%%%%%%%%%%%%%%%%%%%%%%%%%%%%%%%%%%%%%%%%%%%%%%%%%%
In the above, the orbit-averaged precession rates are
%%%%%%%%%%%%%%%%%%%%%%%%%%%%%%%%%%%%%%%%%%%%%%%%%%%%%%%%%%%%%%%%%%%%%%
\be
\Omega_\mathrm{L_\IN S_3}=\frac{GS_3}{2c^2a_\OUT^3(1-e_\OUT^2)^{3/2}}=\Omega_{\mathrm{S_3 L_\IN}}\frac{L_\OUT}{L_\IN}.
\ee
%%%%%%%%%%%%%%%%%%%%%%%%%%%%%%%%%%%%%%%%%%%%%%%%%%%%%%%%%%%%%%%%%%%%%%

\section{B: Modified Evolution of $\hat {\textbf{L}}_\IN$ due to SMBH Spin}
\label{Appendix B}

\begin{figure*}
\includegraphics[width=17cm]{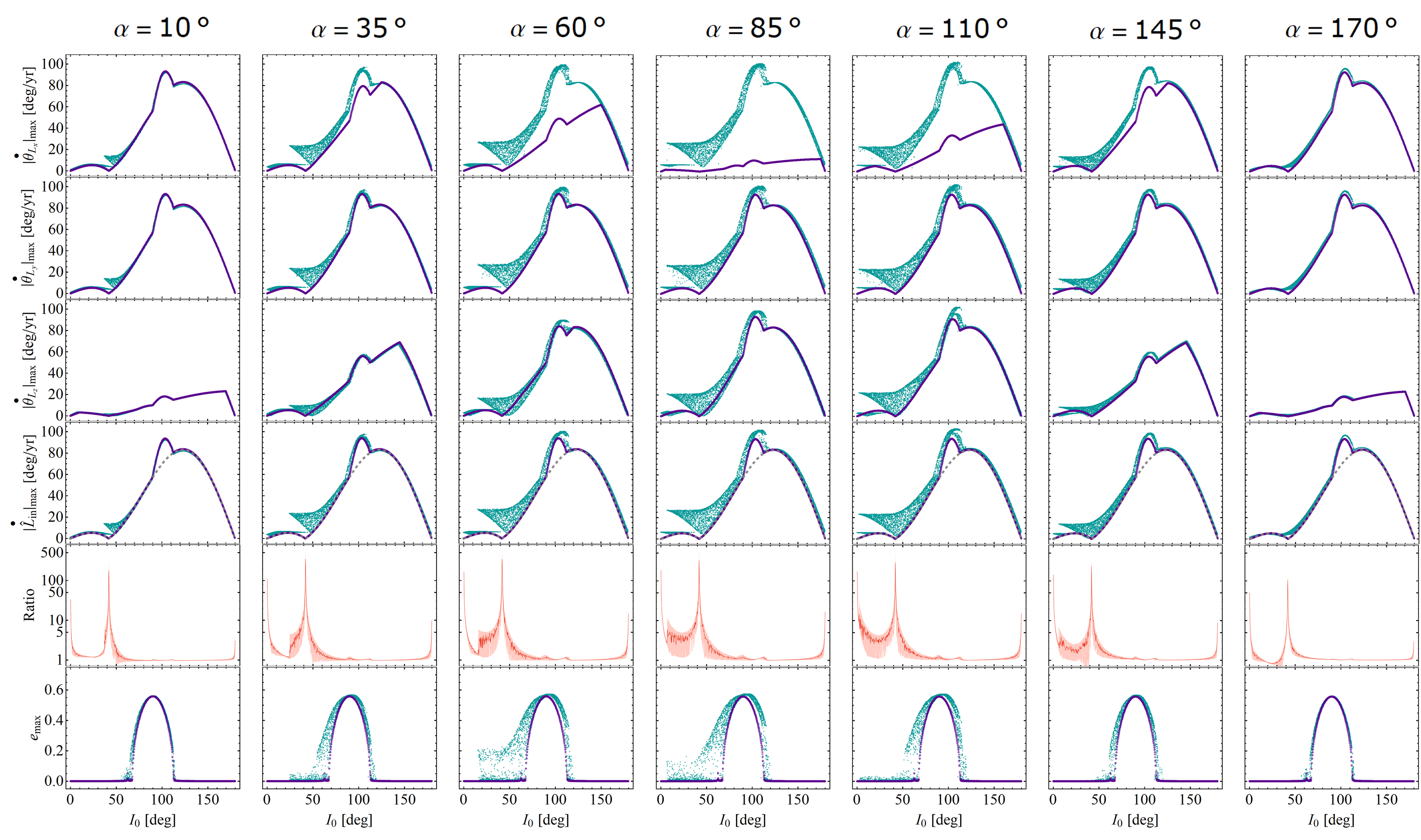}
\caption{Similar to Fig. 4 in the main text, but we consider various values of $\alpha$ (as indicated).
The fifth row shows the ratio $|\dot{\hat {\textbf{L}}}_\IN|_\m(\chi_3=1)/|\dot{\hat {\textbf{L}}}_\IN|_\m(\chi_3=0)$ as
a function of $I_0$, where the shaded region is obtained from the distribution of $|\dot{\hat {\textbf{L}}}_\IN|_\m(\chi_3=1)$
shown in the forth row, and the solid line corresponds to the mean value.
}
\label{fig:Spin window2}
\end{figure*}

\begin{figure*}
\includegraphics[width=17cm]{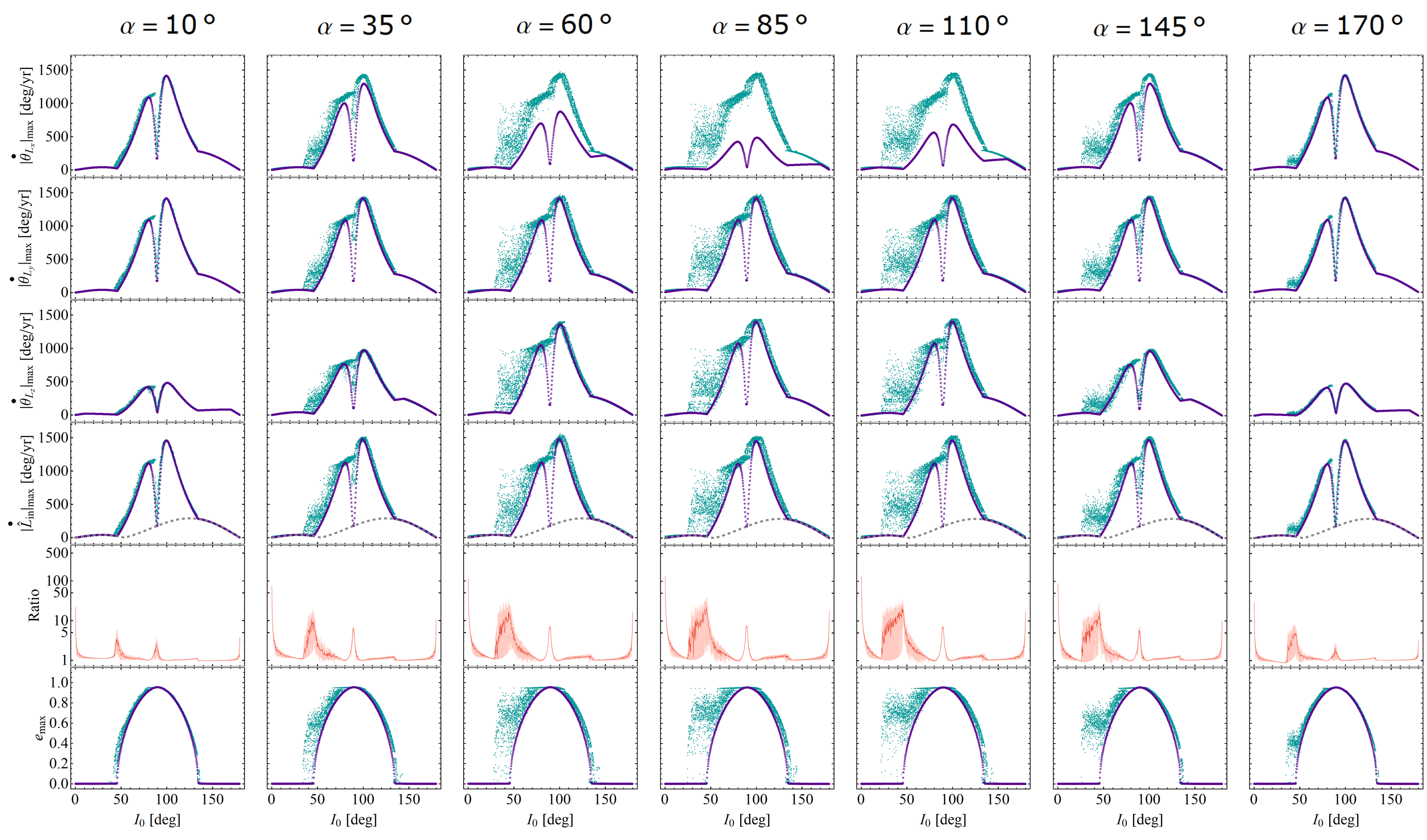}
\caption{Similar to Fig. \ref{fig:Spin window2}, but for the outer binary with $a_\OUT=10\au$.
}
\label{fig:Spin window3}
\end{figure*}

\begin{figure*}
\includegraphics[width=13cm]{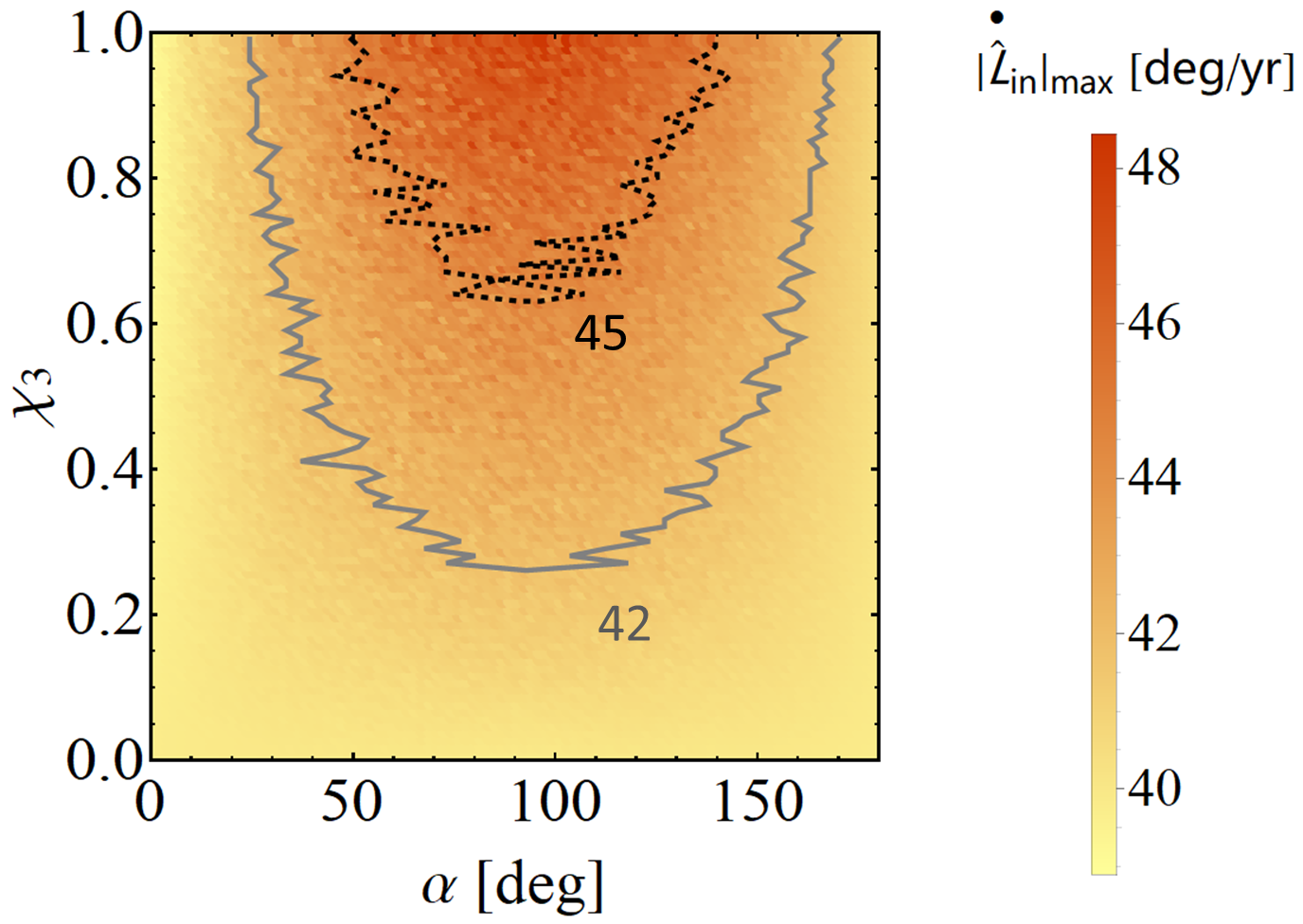}
\caption{Values of the maximum rate of change of $|{\hat {\textbf{L}}}_\IN|$ in $\chi_3-\alpha$ plane.
The system parameters are from the example in Fig. 2 (A) in the main text.
For each combination of ($\chi_3$, $\alpha$), we carry out 100 integrations with a uniform distribution of $I_0$,
and plot the median values of $|d{\hat {\textbf{L}}}_\IN/dt|_\m$.
The two lines (gray dashed and black solid) specify $|d{\hat {\textbf{L}}}_\IN/dt|_\m=42^\circ/\mathrm{yr}, 45^\circ/\mathrm{yr}$, respectively.
}
\label{fig:Spin window4}
\end{figure*}

We explore the effect of the SMBH spin on the variation of orbital axis $\hat {\textbf{L}}_\IN$ and eccentricity $e_\IN$,
taking into account the full range of misalignment angle $\alpha$ (and $\chi_3$). The results are shown in Figs. \ref{fig:Spin window2},
\ref{fig:Spin window3} and \ref{fig:Spin window4}.

\section{C: Signal-to-Noise Ratio}
\label{Appendix C}

An individual binary generates a GW strain composed of discrete harmonics
%%%%%%%%%%%%%%%%%%%%%%%%%%%%%%%%%%%%%%%%%%%%%%%%%%%%%%%%%%%%%%%%%%%%%%
\be
h(t)=\sum\limits_{n=1}^{\infty}h_n(f_n)e^{i 2\pi f_n t},
\ee
%%%%%%%%%%%%%%%%%%%%%%%%%%%%%%%%%%%%%%%%%%%%%%%%%%%%%%%%%%%%%%%%%%%%%%
where $n$ is the number of the harmonics.
The frequency harmonic is
$f_n=n f_\mathrm{orb}$, $f_\mathrm{orb}\equiv\sqrt{Gm_{12}/a^3}/2\pi$, and
%%%%%%%%%%%%%%%%%%%%%%%%%%%%%%%%%%%%%%%%%%%%%%%%%%%%%%%%%%%%%%%%%%%%%%
\be
h_n(f_n)=\frac{2}{n}\sqrt{g(n,e_\IN)}h_0,
\ee
%%%%%%%%%%%%%%%%%%%%%%%%%%%%%%%%%%%%%%%%%%%%%%%%%%%%%%%%%%%%%%%%%%%%%%
here, the function $g(n,e_\IN)$ is given by
%%%%%%%%%%%%%%%%%%%%%%%%%%%%%%%%%%%%%%%%%%%%%%%%%%%%%%%%%%%%%%%%%%%%%%
\be
\begin{split}
&g(n,e_\IN)=\frac{n^4}{32}\Bigg[\Bigg(J_{n-2}-2e_\IN J_{n-1}+\frac{2}{n}J_n+2e_\IN J_{n+1}\\
&-J_{n+1}\Bigg)^2+(1-e_\IN^2)(J_{n-2}-2J_n+J_{n+2})^2+\frac{4}{3n^2}J_n^2\Bigg]
\end{split}
\ee
%%%%%%%%%%%%%%%%%%%%%%%%%%%%%%%%%%%%%%%%%%%%%%%%%%%%%%%%%%%%%%%%%%%%%%
with $J_n\equiv J_n(x)$ is the $i$th Bessel function evaluated at $x=ne_\IN$.
Note that we have introduced the root-mean-square (rms) strain amplitude for the circular orbit at distance $D$
%%%%%%%%%%%%%%%%%%%%%%%%%%%%%%%%%%%%%%%%%%%%%%%%%%%%%%%%%%%%%%%%%%%%%%
\be\label{eq:h 0}
h_0=\sqrt{\frac{32}{5}}\frac{G^2}{c^4}\frac{m_1m_2}{D a_\IN}.
\ee
%%%%%%%%%%%%%%%%%%%%%%%%%%%%%%%%%%%%%%%%%%%%%%%%%%%%%%%%%%%%%%%%%%%%%%
The prefactor $\sqrt{32/5}$ accounts for rms averaging the GW strain over inclination.
Since we only consider the BHB mergers in our MW, Equation (\ref{eq:h 0}) has no redshift dependence.

The signal-to-noise ratio (SNR) is evaluated by
%%%%%%%%%%%%%%%%%%%%%%%%%%%%%%%%%%%%%%%%%%%%%%%%%%%%%%%%%%%%%%%%%%%%%%
\be\label{eq:SNR}
\langle S/N\rangle^2\equiv\int^\infty_0\frac{4|\tilde{h}(f)|^2}{S_n(f)}df
\ee
%%%%%%%%%%%%%%%%%%%%%%%%%%%%%%%%%%%%%%%%%%%%%%%%%%%%%%%%%%%%%%%%%%%%%%
here
%%%%%%%%%%%%%%%%%%%%%%%%%%%%%%%%%%%%%%%%%%%%%%%%%%%%%%%%%%%%%%%%%%%%%%
\be
\tilde{h}(f)=\sum\limits_{n=1}^{\infty}h_n(f_n)T_\mathrm{obs}~\mathrm{sinc}[\pi(f-f_n)T_\mathrm{obs}],
\ee
%%%%%%%%%%%%%%%%%%%%%%%%%%%%%%%%%%%%%%%%%%%%%%%%%%%%%%%%%%%%%%%%%%%%%%
$T_\mathrm{obs}$ is the observation time,
and $S_n(f)$ is the full strain spectral sensitivity density.
Here, we consider the LISA instrumental noise and confusion noise from the unresolved galactic binaries \citep[e.g.,][]{LISA curve}.

If the system has the orbital decay timescale much longer than the LISA mission time,
we have
%%%%%%%%%%%%%%%%%%%%%%%%%%%%%%%%%%%%%%%%%%%%%%%%%%%%%%%%%%%%%%%%%%%%%%
\be\label{eq:SNR 2}
\langle S/N\rangle^2=\sum\limits_{n=1}^{\infty}\frac{4|h_n(f_n)|^2(f_nT_\mathrm{obs})}{f_nS_n(f_n)}.
\ee
%%%%%%%%%%%%%%%%%%%%%%%%%%%%%%%%%%%%%%%%%%%%%%%%%%%%%%%%%%%%%%%%%%%%%%
Also, we can define the characteristic strain as
%%%%%%%%%%%%%%%%%%%%%%%%%%%%%%%%%%%%%%%%%%%%%%%%%%%%%%%%%%%%%%%%%%%%%%
\be\label{eq:Characteristic Strain 2}
h_{c,n}=2h_n(f_n)\sqrt{f_n T_\mathrm{obs}}.
\ee
%%%%%%%%%%%%%%%%%%%%%%%%%%%%%%%%%%%%%%%%%%%%%%%%%%%%%%%%%%%%%%%%%%%%%%
Equations (\ref{eq:SNR 2})-(\ref{eq:Characteristic Strain 2}) suggest that SNR can be enhanced by a factor of $\sqrt{f_n T_\mathrm{obs}}$.

\begin{figure*}
\includegraphics[width=10cm]{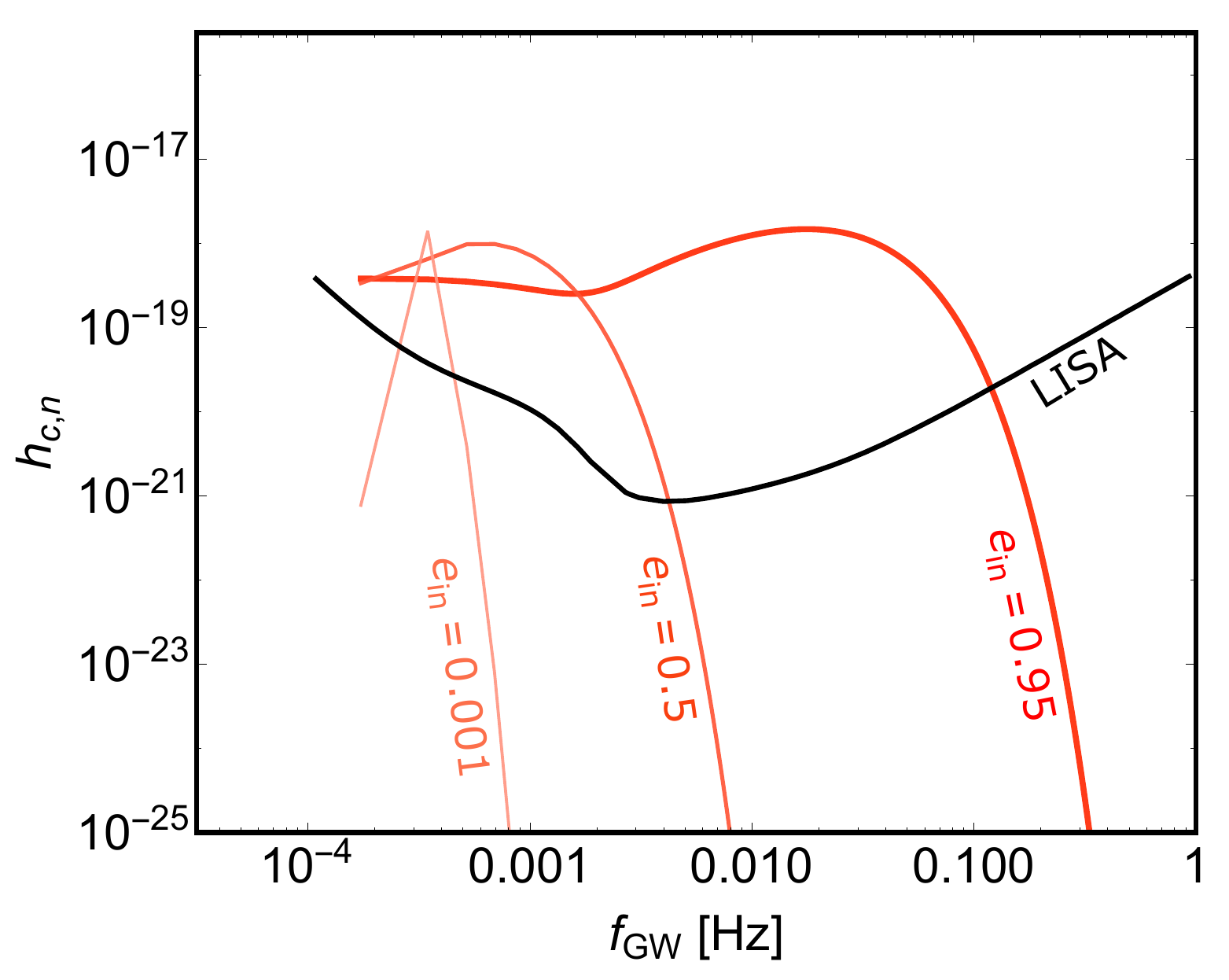}
\caption{The GW strain curve of BH binary ($m_1=20M_\odot$, $m_2=10M_\odot$ and $a_\IN=0.01$AU).
Different orbital eccentricities are taken into account (as labeled).
The red curves are obtained by Eq. (\ref{eq:Characteristic Strain 2}) with a series of $n$, and
the spectral sensitivity density of LISA is from \citet{LISA curve}.
}
\label{fig:SNR}
\end{figure*}

We find that for the GW sources ($m_1=20M_\odot$, $m_2=10M_\odot$ and $a_\IN=0.01$AU),
the total SNR is greater than 37 if we assume the observation time $T_\mathrm{obs}=5$yrs and the distance $D=8$kpc.
An elliptic inner orbit can enhance the detectability, increasing the overall SNR by a factor of
10 (or even more), as shown in Fig. \ref{fig:SNR}.

\section{D: Modified GW Waveforms}
\label{Appendix D}

As the compact binary precesses, both amplitude and phase of the GW waveform can be modified.
The signature of the time-varying orientation $\hat{\textbf{L}}_\IN$ can be extracted through
projecting the GW radiation onto the antenna (detector) coordinates \citep[e.g.,][]{Yuhang PRL}.

Considering the circular inner orbit, the waveform is expressed in terms of frequency
%%%%%%%%%%%%%%%%%%%%%%%%%%%%%%%%%%%%%%%%%%%%%%%%%%%%%%%%%%%%%%%%%%%%%%
\be
\tilde{W}(f)=\Lambda(f)\tilde{W}_C(f),
\ee
%%%%%%%%%%%%%%%%%%%%%%%%%%%%%%%%%%%%%%%%%%%%%%%%%%%%%%%%%%%%%%%%%%%%%%
where $\tilde{W}_C(f)$ is the antenna-independent ``carrier", which is a function of the chirp mass, distance, time and phase of coalescence, and
%%%%%%%%%%%%%%%%%%%%%%%%%%%%%%%%%%%%%%%%%%%%%%%%%%%%%%%%%%%%%%%%%%%%%%
\be
\begin{split}
\Lambda(f)\equiv&\sqrt{A_+^2(t)F_+^2(t)+A_\times^2(t)F_\times^2(t)}\\
&\times \mathrm{exp}\big\{-i[\Phi_p(t)+2\Phi_T(t)+\Phi_D(t)]\big\}.
\end{split}
\ee
%%%%%%%%%%%%%%%%%%%%%%%%%%%%%%%%%%%%%%%%%%%%%%%%%%%%%%%%%%%%%%%%%%%%%%
Here, we have introduced the amplitude terms
%%%%%%%%%%%%%%%%%%%%%%%%%%%%%%%%%%%%%%%%%%%%%%%%%%%%%%%%%%%%%%%%%%%%%%
\ba
&&A_+(t)=1+(\hat{\textbf{L}}_\IN(t)\cdot \hat{\textbf{N}})^2,\\
&&A_\times(t)=-2\hat{\textbf{L}}_\IN(t)\cdot \hat{\textbf{N}},
\ea
%%%%%%%%%%%%%%%%%%%%%%%%%%%%%%%%%%%%%%%%%%%%%%%%%%%%%%%%%%%%%%%%%%%%%%
where $\hat{\textbf{N}}$ is the direction of line-of-sight,
and $F_{+(\times)}(t)$ is the antenna pattern coefficient. For the
phase terms, $\Phi_p$ shows the polarization phase
%%%%%%%%%%%%%%%%%%%%%%%%%%%%%%%%%%%%%%%%%%%%%%%%%%%%%%%%%%%%%%%%%%%%%%
\be
\Phi_p(t)=\arctan \bigg[-\frac{A_\times(t)F_\times(t)}{A_+(t)F_+(t)}\bigg],
\ee
%%%%%%%%%%%%%%%%%%%%%%%%%%%%%%%%%%%%%%%%%%%%%%%%%%%%%%%%%%%%%%%%%%%%%%
$\Phi_T$ characterizes the Thomas precession
%%%%%%%%%%%%%%%%%%%%%%%%%%%%%%%%%%%%%%%%%%%%%%%%%%%%%%%%%%%%%%%%%%%%%%
\be
\Phi_T(t)=-\int dt
\bigg[\frac{\hat{\textbf{L}}_\IN(t)\cdot \hat{\textbf{N}}}{1-(\hat{\textbf{L}}_\IN(t)\cdot \hat{\textbf{N}})^2}\bigg]
(\hat{\textbf{L}}_\IN(t)\times \hat{\textbf{N}})\cdot\frac{d\hat{\textbf{L}}_\IN(t)}{dt},
\ee
%%%%%%%%%%%%%%%%%%%%%%%%%%%%%%%%%%%%%%%%%%%%%%%%%%%%%%%%%%%%%%%%%%%%%%
and $\Phi_D$ is the Doppler phase induced by the motions of outer orbit (and/or the detector orbiting around the Sun).

Therefore, the change of the $\hat{\textbf{L}}_\IN$ orientation can be tracked for the GW source with sufficient SNR.
\citet{Yuhang PRL} explored the detectability of the precessing $\hat{\textbf{L}}_\IN$ around $\hat{\textbf{L}}_\OUT$.
In their study, the Newtonian precession and the de-Sitter like precession (Effect II here) are taken into account,
and the evolution of $\hat{\textbf{L}}_\IN$
is regular. It is suggested that the orientation change is measurable if the precession timescale is less than the observation time.
Similarly, since the spin-induced precession timescale we considered is comparable or less than 10 yrs,
the frame-dragging effect in the orbital plane might also be
detected through the modified GW waveform---the only thing is $\hat{\textbf{L}}_\IN$ evolves in a non-regular way.

%%%%%%%%%%%%%%%%%%%%%%%%%%%%%%%%%%%%%%%%%%%%%%%%%%%%%%%%%%%%%%%%%%%%%%

\end{document}